\documentclass[twocolumn,preprintnumbers,aps, pra,10pt,floatfix]{revtex4-2}

\usepackage{graphicx,amsmath}
\usepackage{dcolumn}
\usepackage{bm}
\usepackage{amssymb}
\usepackage{color}
\graphicspath{{Figures/}}

\begin{document}

 \title{Quantum Rabi oscillations of a qubit strongly coupled to
 a one-dimensional waveguide}

\begin{abstract}

We theoretically investigate quantum Rabi oscillations in a system
consisting of a two-level atom (qubit) strongly coupled to a
one-dimensional open waveguide. In contrast to conventional cavity
quantum electrodynamics, the qubit interacts with a continuum of
propagating modes, which gives rise to fundamentally different
dynamical behavior. Within the rotating-wave approximation, we
express the multimode Jaynes-Cummings Hamiltonian in terms of
collective bosonic operators and show that the system possesses
two integrals of motion, enabling an exact diagonalization of the
Hamiltonian in the single-excitation subspace. We show that for a
strongly interacting qubit-photon system, the dynamics are
captured by a reduced two-level model. In this framework, each
level is defined as the product of the atomic excited state and a
specific field mode, which takes the form of a multiphoton
Fock-like state. In this picture, the Rabi oscillations represent
a collective phenomenon corresponding to oscillations between
multiphoton states differing by a single photon. We then extend
our analysis to multiphoton processes in which the initial field
is a coherent state with a continuous spectrum. In this case, the
Rabi frequency is shown to be sensitive to the spectral profile of
the function that generates the coherent state.

\end{abstract}


\date{\today}

\author{Ya. S. Greenberg}\email{yakovgreenberg@yahoo.com}
\affiliation{Department of Applied and Theoretical Physics,
Novosibirsk State Technical University, Novosibirsk 630073,
Russia}

\author{A. A. Shtygashev} \affiliation{Department of Applied and Theoretical Physics,
Novosibirsk State Technical University, Novosibirsk 630073,
Russia}

\author{O. A. Chuikin} \affiliation{Department of Applied and Theoretical Physics,
Novosibirsk State Technical University, Novosibirsk 630073,
Russia}

\author{A. G. Moiseev} \affiliation{Department of Applied and
Theoretical Physics, Novosibirsk State Technical University,
Novosibirsk 630073, Russia}

\author{O. V. Kibis} \affiliation{Department of Applied and Theoretical Physics,
Novosibirsk State Technical University, Novosibirsk 630073,
Russia}

\maketitle

\section{Introduction}

Quantum Rabi oscillations in the regime of strong coupling between
an atom and an electromagnetic field are a coherent reversible
process in which a photon is periodically absorbed and emitted by
the atom at a frequency equal to the coupling energy divided by
Planck's constant \cite{Brune1996, Raim2001}. For many years,
cavity quantum electrodynamics (cQED) has been the primary
platform for studying strong-coupling effects and quantum Rabi
oscillations in quantum optics and solid-state nanostructures with
artificial atoms (qubits) \cite{Brune2004, Blais2004, Wall2004,
Wal2006, Blais2021}.

The progress of quantum technologies in recent years has made it
possible to develop an experimental platform alternative to cQED.
This refers to waveguide quantum electrodynamics (wQED), in which
an atom or qubit interacts with a continuum of electromagnetic
field modes in an open waveguide or coplanar open strip line
\cite{Roy2017, Sher2023, Gu2017}.

Quantum Rabi oscillations in an open waveguide represent a
fundamental quantum phenomenon in wQED, where a quantum emitter
(such as an atom, quantum dot, or superconducting qubit)
coherently exchanges energy with the vacuum field of an open
waveguide which supports a continuum of propagating modes rather
than discrete cavity modes leading to different density of states
characteristics (see \cite{Mukh2024} where the extensive review of
the relevant literature was provided). This phenomenon occurs in
the strong coupling regime, where the emitter-waveguide coupling
rate exceeds both the emitter decay rate and the waveguide loss
rate.

It is generally believed that Rabi oscillations in an open system
can exist only in the ultra-strong coupling regime, when the
qubit-electromagnetic-field interaction and the Lamb shift are
comparable to the qubit excitation frequency \cite{Cohen2004}.
However, such oscillations are also possible under conditions
where the coupling strength is much smaller than the qubit
frequency but much larger than the qubit damping rate. As
calculations show, such oscillations are possible in
one-dimensional multimode waveguide structures such as an open
waveguide \cite{Mukh2024, Green2025} or a broadband cavity
\cite{Green2026}.

Unlike quantum Rabi oscillations in a cavity, the experimental
observation of Rabi oscillations in an open waveguide presents a
serious challenge \cite{Sult2025}. Quantum Rabi oscillations in
open waveguide occur due to the interaction with vacuum
fluctuations and multimode electromagnetic field, making them a
purely quantum phenomenon. In open waveguide, these oscillations
are significantly modified by the continuous spectrum of modes
and/or the non-orthogonal nature of waveguide eigenmodes
\cite{Mukh2024,Cheng2008}.

For  Rabi oscillations (i.e., oscillations in the qubit excitation
amplitude) to be observed in an open waveguide, the damping rate
of these oscillations $\Gamma_q$ must be much smaller than the
Rabi frequency $\Omega_R$, i.e., $\Gamma_q \ll \Omega_R$. This
condition seems difficult to satisfy, since both $\Omega_R$ and
$\Gamma_q$ depend on the same parameter-the coupling energy.
However, the specific details of the qubit's behavior depend to a
large extent on the density of states of the electromagnetic field
spectrum. Rabi oscillations originate from the coupling of the
qubit to the frequency components of the continuum close to the
qubit transition frequency, while the decay of these oscillations
is governed by the qubit's interaction with continuous field modes
far detuned from the qubit excitation frequency.

The problem of calculating Rabi oscillations for the single-mode
Jaynes-Cummings Hamiltonian can be solved exactly \cite{Scul1997}.
However, extending this problem to the case of several discrete
modes or a continuous spectrum encounters significant difficulties
\cite{Swain1972a, Swain1972b, David1974, Mukh2024}. Even though
the evolution of the excited-state amplitude of the qubit can be
written exactly as a contour integral of the Green's function
\cite{David1974}, the evaluation of this integral is so
complicated that the main approach to solving this problem is to
choose a suitable trial wave function, which in a certain
parameter regime yields the correct result.

In the present work, the trial wave function describing a quantum
superposition of the qubit state and the state of a multiphoton
continuous spectrum is chosen such that the obtained result is a
natural generalization of the single-mode problem. The
qubit-waveguide electromagnetic field interaction is taken to be
in the strong-coupling regime, which, following the standard
classification, implies that the coupling strength $g/\hbar$ is
significantly less than the qubit transition frequency $\Omega$,
yet significantly greater than the relaxation rate $\Gamma_q$.
This justifies the neglect of counter-rotating terms in the
qubit-field interaction Hamiltonian. The trial wavefunction  we
have chosen does not allow us to calculate the intrinsic damping
of the qubit. Therefore, in the present work, we do not compute
the decay rate of the qubit excitation amplitude $\Gamma_q$ and
assume that the condition $\Gamma_q \ll \Omega_R$ is satisfied.

The paper is organized as follows. In Section II, we introduce the
Hamiltonian for the qubit-field system. The coupling between the
qubit and the waveguide photon modes is described by a multimode
Jaynes-Cummings Hamiltonian in the rotating-wave approximation. It
is shown that the total Hamiltonian can be expressed in terms of
two integrals of motion, enabling an exact diagonalization of the
Hamiltonian in the single-excitation subspace. Section III is
dedicated to analyzing the wave-function evolution of the
qubit-field system for the initial condition where the qubit is in
its excited state and the waveguide is in the photon vacuum state.
By making a suitable ansatz for the trial wave function, we are
able to study the dynamics of the amplitude of the initially
excited qubit, which undergoes Rabi-like oscillations at a
frequency, determined by the effective qubit-field coupling in the
waveguide. This dynamics corresponds to a periodic exchange of a
single photon between the qubit and a superposition of a continuum
multimode collective excitations, with each mode carrying one
photon. For Lorentzian density of states of photon modes, we find
that the oscillation frequency and the frequency detuning are
strongly influenced by the damping rate of the propagating photon
modes in the waveguide. In Section IV, we define the multiphoton
continuum Fock number states, which are then used in the later
sections to build up the multiphoton \(A\)-number states and
\(a_{\alpha}\)-number states.  In Section V \(A\)-number states
are used to build a trial wavefunction where initially qubit is
excited and there exists $n-1$ photons which are distributed over
all frequencies in a multiphoton \(A\) state. In this case, Rabi
oscillations represent a periodic exchange of a single photon
between the qubit and a superposition of multiphoton collective
excitations, with each mode carrying arbitrary number of photons,
$n_{\omega}$. For this case, the Rabi frequency is shown to be
proportional to the product of the effective coupling strength
$\sqrt{\Lambda}$ and $\sqrt{n}$. In Section VI, we investigate the
evolution of the wave function when the qubit is initially in the
excited state and the initial field is a continuum coherent state.
In this case, the time dependence of the wave-function amplitude
exhibits an oscillatory process involving the exchange of a photon
between multiphoton continuum states differing by one photon.
Moreover, in contrast to the single-mode case, the frequency of
these oscillations depends significantly on the spectral shape of
the function $\alpha(\omega)$ that generates the coherent state.
The Rabi frequency is calculated for a Gaussian form of
$\alpha(\omega)$, with the photon density of states taken as a
Lorentzian. It is also shown that the Rabi frequency depends
significantly on the loss rate of the photon mode.

\section{Theoretical model}

We consider a system consisting of a two-level atom (qubit),
interacting with the photonic modes of a one-dimensional open
waveguide. The qubit is placed at the origin, \(x = 0\). The
interaction between the qubit and the electromagnetic field of the
waveguide is described by multimode Jaynes-Cummings Hamiltonian in
the rotating-wave approximation (RWA). This implies that the
coupling strength \(g\) between the qubit and the field is much
smaller than the qubit excitation frequency \(\Omega\).

Therefore, we start from the Hamiltonian:
   \begin{equation}\label{1}
H = H_0  + H_{JC},
   \end{equation}
where:
   \begin{equation}\label{2}
H_0  = \Omega \sigma _ +  \sigma _ -   + \sum\limits_k {\omega_k
a_k^\dagger a_k},
   \end{equation}
   \begin{equation}\label{HJC}
H_{JC}  = \sum\limits_k {(g_k a_k^\dagger \sigma_-  }  + g_k^*
a_k \sigma_+ )
   \end{equation}
In (\ref{2}) $\Omega$ is a resonant frequency of a qubit,
$\sigma_- = {\left|g\right\rangle} \left\langle e \right|$ and
$\sigma _ +  = {\left| e \right\rangle } \left\langle g \right|$
are the lowering and raising atomic operators which lower or raise
a state of a qubit, where $|e\rangle$ and $|g\rangle$ are excited
and ground states of a two-level atom,  $\omega$ is a photon
frequency. The photon creation and annihilation operators
$a^\dagger_k$, $a_k$,  satisfy the boson commutation relation:
   \begin{equation}\label{comut}
[a_k,a^\dagger_{k'}]= \delta_{k,{k'}}.
  \end{equation}

In (\ref{HJC}) $H_{JC}$ is the  continuous mode Jaynes-Cummings
Hamiltonian which describes the interaction of the photon field
with a qubit located at the point $x=0$. The quantity $g_k$ is the
coupling between qubit and the photon field in a waveguide.

Throughout the paper we set $\hbar= 1$ so that all energies are
expressed in frequency units. Therefore, the dimension of the
coupling constant $g_k$ is a frequency, ${\omega}$.

In what follows, it is convenient to rewrite Hamiltonian
(\ref{HJC}) in terms of collective operators $A^\dagger, A$ which
satisfy the boson commutation relation:
   \begin{equation}\label{Com}
    [A,A^\dagger]=1,
   \end{equation}
where:
   \begin{equation}\label{4}
A^\dagger   = \frac{1}
{{\sqrt \Lambda  }}\sum\limits_k {g_k a_k^\dagger  } , \qquad
A = \frac{1}
{{\sqrt \Lambda  }} \sum\limits_k {g_k^* a_k },
   \end{equation}
and:
   \begin{equation}\label{Lam}
\Lambda  = \sum\limits_k {\left| {g_k } \right|^2 },
   \end{equation}
is the square of the effective coupling strength between qubit and
the photon modes in a waveguide.
Therefore, Hamiltonian (\ref{HJC}) takes the form:
   \begin{equation}\label{3}
H_{JC}  = \sqrt{\Lambda}A^\dagger  \sigma_-   + \sqrt{\Lambda}A \sigma
_ +.
   \end{equation}

It is not difficult to see that Hamiltonian (\ref{1}) allows for
two integrals of motion:
   \begin{equation}\label{5}
\widehat{N} = \sigma_ +  \sigma_-  + \sum\limits_k {a_k^\dagger  a_k },
   \end{equation}
   \begin{equation}\label{6}
\widehat{Q} = \sum\limits_k {} (\omega _k  - \Omega )a_k^\dagger  a_k  + H_{JC},
   \end{equation}
where $[\widehat{Q},H]=0$, $[\widehat{N},H]=0$.

Hamiltonian (\ref{1}) can be expressed in terms of these operators
as follows:
   \begin{equation}\label{7}
H = \widehat{Q} + \widehat{N}\Omega.
   \end{equation}
Therefore, any wavefunction $\Psi$, which is a solution of the
Schr\"{o}dinger equation $H\Psi=E\Psi$, must be eigenfunction of
aforementioned operators $\widehat{N}$ and $\widehat{Q}$ with
eigenvalues $n$ and $q$, correspondingly: $\widehat{N}\Psi=n\Psi$,
$\widehat{Q}\Psi=q\Psi$. Thus, from (\ref{7}), the energy of
the system can be written as $E=q+n\Omega$. The integer positive
quantity $n$ describes the number of excitation in a system. A
system consisting of an excited qubit and $(n - 1)$ photons has the
same total number of excitations ($n$) as a system with a qubit in
the ground state and $n$ photons.

We note that as $\widehat{N}$ commutes with $H$, the system
initially prepared in an eigenstate of $\widehat{N}$ having
eigenvalue $n$, will remain in an eigenstate of $\widehat{N}$
belonging to the same eigenvalue $n$ for all time. Thus a complete
set of states for describing a single atom placed in its excited
state in a cavity at $t = 0$ with no photons present initially is
obtained by listing all the eigenstates of $\widehat{Q}$ belonging
to the eigenvalue $n$.

\section{Single-photon Rabi oscillations}

The task is to find the time-dependent state vector $\left| {\Psi
(t)} \right\rangle$ for the initial state $\left| {\Psi (0)}
\right\rangle = \left| {e,0} \right\rangle$. As is well known, the
exact solution of this problem is given by the evolution operator
$U(t)$, $ \left| {\Psi (t)} \right\rangle  = U(t)\left| {\Psi (0)}
\right\rangle $, where in our case $ U(t) = \exp(-i\widehat{Q}t)$.
For strong interaction we assume that in equation (\ref{6}) the
leading term is $H_{JC}$. Therefore, we may write
$|\Psi(t)\rangle\approx \exp({-iH_{JC}t})|e,0\rangle$. The direct
calculation provides the following expression (see Appendix
\ref{A}):
   \begin{equation}\label{operU}
\begin{gathered}
    |\Psi(t)\rangle\approx \exp({-iH_{JC}t}) |e,0\rangle
    \\
    = \cos\left( \sqrt{\Lambda} t \right) |e,0\rangle
    -i\sin\left(\sqrt{\Lambda}t\right)A^\dagger |g,0\rangle.
    \end{gathered}
   \end{equation}

As was shown in \cite{Green2025} this expression yields two Rabi
peaks in the spectrum of the first-order correlation function of
the electric field.

Therefore, it seems reasonable to choose a trial stationary
wavefunction from a single-excitation subspace in the following
form:
   \begin{equation}\label{8}
\Psi _1  = A_1 \left| {e,0} \right\rangle  + B_1 A^\dagger  \left|
{g,0} \right\rangle,
   \end{equation}
with the normalization
   \begin{equation}\label{norm}
\left| {A_1 } \right|^2  + \left| {B_1 } \right|^2  = 1.
   \end{equation}

 It is not difficult to show that
$\widehat{N}\Psi_1=\Psi_1$. The eigenvalue $q_1$ of the operator
$\widehat{Q}$ can be found from the solution of the Schr\"odinger
equation $\widehat{Q}\Psi_1=q_1\Psi_1$:
   \begin{equation}\label{17}
\begin{gathered}
  B_1 \sqrt \Lambda   = q_1 A_1,
  \\
  A_1 \sqrt \Lambda   + B_1 G = q_1 B_1,
\end{gathered}
   \end{equation}
where:
   \begin{equation}\label{18}
G = \frac{1} {\Lambda }\sum\limits_k {} (\omega _k  - \Omega
)\left| {g_k } \right|^2,
   \end{equation}
   \begin{equation}\label{19}
q_{1, \pm }  = \frac{G} {2} \pm \frac{1} {2}\sqrt {G^2  + 4\Lambda
}.
   \end{equation}

The quantity $G$ can be transformed to:
   \begin{equation}\label{20}
G = \bar \omega  - \Omega,
   \end{equation}
where $\bar{\omega}$ is the weighted average frequency:
   \begin{equation}\label{21}
\bar \omega  = \sum\limits_k \omega_k \left| {g_k } \right|^2.
   \end{equation}

Therefore, there are two wavefunctions $\Psi_{1,+}$ and
$\Psi_{1,-}$ corresponding to $q_{1,+}$ and $q_{1,-}$:
   \begin{equation}\label{22}
\begin{gathered}
  \Psi _{1 + }  = A_{1 + } \left| {e,0} \right\rangle  + B_{1 + } A^ +  \left| {g,0} \right\rangle,
  \\
  \Psi _{1 - }  = A_{1 - } \left| {e,0} \right\rangle  + B_{1 - } A^ +  \left| {g,0} \right\rangle .
\end{gathered}
   \end{equation}

Knowing $q_{1,\pm}$ and normalization (\ref{norm}) we can find
from (\ref{17}) the coefficients $A_{1,\pm}$ and $B_{1,\pm}$:
   \begin{equation}\label{23}
\begin{gathered}
  A_{1 + }  = \cos \theta ,\;B_{1 + }  = \sin \theta,
  \\
  A_{1 - }  = \sin \theta ,\;B_{1 - }  =  - \cos \theta,
\end{gathered}
   \end{equation}
where:
   \begin{equation}\label{24}
\tan\theta  = \frac{{q_{1 + } }} {{\sqrt \Lambda  }} =  -
\frac{{\sqrt \Lambda  }} {{q_{1 - } }},
   \end{equation}
   \begin{equation}\label{25}
\sin \theta  = \frac{{q_{1 + } }} {{\sqrt {\Lambda  + q_{1 + }^2 }
}};\quad \cos\theta  = \frac{\Lambda } {{\sqrt {\Lambda  + q_{1 +
}^2 } }}.
   \end{equation}

Now it is not difficult to construct the time-dependent
wavefunction $\Psi(t)$ with the initial condition
$\Psi(0)=|e,0\rangle$:
   \begin{equation}\label{26}
\begin{gathered}
\Psi (t) = \left( {e^{ - iE_{1+}  t} \cos ^2 \theta  + e^{ - iE_{1 - } t} \sin^2 \theta } \right)\left| {e,0} \right\rangle
\\
+ \cos \theta \sin \theta \left( {e^{ - iE_{1+}  t}  - e^{ - iE_{1 - } t} } \right)A^\dagger  \left| {g,0} \right\rangle,
\end{gathered}
   \end{equation}
where $ E_{1 \pm }  = q_{1 \pm }  + \Omega$.

It is worth noting that in the limit $\sqrt{\Lambda}\gg G$, one
has $\cos\theta=\sin\theta=\sqrt{1/2}$ and, consequently, eq.
(\ref{26}) reduces exactly to eq. (\ref{operU}).

Thus, for a qubit initially in the excited state and a waveguide
initially containing no photons, the probability that the qubit
remains excited at time $t$ is given by:
   \begin{equation}\label{27}
\begin{gathered}
P_e (t) = \left| {e^{ - iE_{1 + } t} \cos ^2 \theta  + e^{ - iE_{1 - } t} \sin ^2 \theta } \right|^2
\\
= 1 - 4\sin ^2 \theta \cos ^2 \theta \sin ^2 \frac{{E_{1 + }  -
E_{1 - } }}
{2}t.
\end{gathered}
   \end{equation}

Alternatively, the probability of finding the qubit in the ground
state with one photon in the waveguide at time $t$ is as follows:
   \begin{equation}\label{28}
\begin{gathered}
  P_g (t) = \cos ^2 \theta \sin ^2 \theta \left| {\left( {e^{ - iE_{1 + } t}  - e^{ - iE_{1 - } t} } \right)} \right|^2
  \\
   = 4\sin ^2 \theta \cos ^2 \theta \sin ^2 \frac{{E_{1 + }  - E_{1 - } }}
{2}t.
\end{gathered}
   \end{equation}

In equations (\ref{27}), (\ref{28}):
   \begin{equation}\label{29}
E_{1 + }  - E_{1 - }  = q_{1 + }  - q_{1 - }  = \sqrt {G^2  +
4\Lambda }.
   \end{equation}

Another important quantity is the inversion $ W(t) = \left\langle
{\Psi (t)} \right|\sigma _Z \left| {\Psi (t)} \right\rangle$,
 which is related to the probabilities $P_e(t)$ and $P_g(t)$
by the expression:
   \begin{equation}\label{30}
\begin{gathered}
W(t) = P_e (t) - P_g (t)\\ = 1 - 8\cos ^2 \theta \sin ^2 \theta
\sin ^2 \frac{{E_ {1+ }  - E_{1 -}  }} {2}t.
\end{gathered}
   \end{equation}

In case of $\Lambda\gg G^2$ we obtain from (\ref{25})
$\sin\theta=\cos\theta=\frac{1}{\sqrt{2}}$ and:
   \begin{equation}\label{31}
   \begin{gathered}
P_e (t) = \cos ^2 \sqrt \Lambda  t, \qquad P_g (t) = \sin ^2 \sqrt
\Lambda  t,
\\
W(t) = \cos 2\sqrt{\Lambda} t.
\end{gathered}
   \end{equation}

From the formal point of view the expressions (\ref{27}),
(\ref{28}), (\ref{30}) are similar to those obtained for a single
mode case \cite{Scul1997}. It is easily seen with the substitution
of $g^2$ and frequency detuning $\Delta$ for $\Lambda$ and $G$ in
these expressions.

However, a crucial distinction exists: in the single-mode state,
the photon oscillates between the vacuum photon state
$|e,0\rangle$ and the single-mode photon state
$|g,1\rangle=|g\rangle\bigotimes a^\dagger_k|0\rangle$. In our case, the
photon oscillates between the vacuum photon state $|e,0\rangle$
and the multi-mode state
$A^\dagger|g,0\rangle=|g\rangle\bigotimes\frac{1}{\sqrt{\Lambda}}\sum_k
g_k a^\dagger_k|0\rangle$, where each one-photon state is weighted by the
frequency-dependent factor $g_k/\sqrt{\Lambda}$. Thus, for a given
weighted frequency $\overline{\omega}$, defined in (\ref{21}), a qubit emits
the photon with any frequency $\omega_k$ with the probability
amplitude $\approx g_k/\sqrt{\Lambda} $, and later the photon
comes back to the emitter with the frequency which can be
different from $\omega_k$. In other words, the probability
amplitudes oscillate between the vacuum state $|e,0\rangle$ and
the ensemble of single-photon states representing the collective
effect of the single-photon Rabi oscillations.

We see that the frequency of Rabi oscillations is equal to
$2\sqrt{\Lambda}$, where $\sqrt{\Lambda}$ represents the effective
coupling strength between qubit and the photon field in a
waveguide. The final results, obtained in this section, depend
only on two quantities, $G$ and $\Lambda$, defined in (\ref{18})
and (\ref{Lam}), respectfully.

In order to transit to the continuous limit, we define the
spectral function
   \begin{equation}\label{J}
J(\omega ) = \sum\limits_k {} \left| {g_k } \right|^2 \delta
(\omega  - \omega _k ),
   \end{equation}
that allows us to rewrite the quantities $\Lambda$, $G$, and
$\bar{\omega}$ as follows:
   \begin{equation}\label{Lam1}
\Lambda  = \int\limits_0^\infty  {} J(\omega )d\omega,
   \end{equation}
   \begin{equation}\label{G1}
G = \frac{1} {\Lambda }\int\limits_0^\infty  {d\omega \;} (\omega
- \Omega )J(\omega ),
   \end{equation}
   \begin{equation}\label{avfr}
\overline{\omega}  =\frac{1}{\Lambda} \int\limits_0^\infty
{d\omega \,\omega J(\omega )}.
   \end{equation}

From the definitions of $\Lambda$ (\ref{Lam}), (\ref{Lam1}) it
follows that in the continuous limit the quantity $J(\omega)$ is a
product of the photon density of states $D(\omega)$ and the
coupling strength $g^2(\omega)$, $J(\omega)=D(\omega)g^2(\omega)$.
Below we define $J(\omega)=\Lambda f(\omega)$, where:

   \begin{equation}\label{ds}
f(\omega ) = \frac{1}{\Lambda}D(\omega)g^2 (\omega ),
   \end{equation}
so that $\int\limits_0^\infty f(\omega )d\omega = 1$. This
definition guarantees that condition (\ref{Lam1}) is satisfied.

We take the photon density of states, $D(\omega)$ in the form of
the Lorentzian line, $D(\omega)$:

   \begin{equation}\label{ds1}
D(\omega ) = \frac{{\Gamma }} {{(\omega  - \omega _0)^2  +
\Gamma^2 }},
   \end{equation}

where $\omega_0$ is the frequency of the fundamental mode (which
is close to the qubit frequency $\Omega$), $1/\Gamma$ is a decay
constant.

The coupling strength $g(\omega)$ in (\ref{ds}) refers to 1D free
space while $g(k)$ in (\ref{J}) represents 3D free space. We
specified this difference in Appendix \ref{B} where we describe
the transition from the summation over discrete quantities to the
integration over continuum variables.

If in (\ref{ds}), $g^2(\omega)$ were assumed to be constant, the
integral (\ref{avfr}) would diverge at the upper bound. In quantum
electrodynamics there are restrictions on possible forms of
$g^2(\omega)$ which prevent the infrared and ultraviolet
divergences. Here we adopt the following form of $g^2(\omega)$
\cite{David1974}:
   \begin{equation}\label{g2}
g^2 (\omega ) =g^2(\Omega)\times \left\{ \begin{gathered}
  \frac{\omega}{\Omega} \quad \quad \omega  \leqslant \Omega
  \\
  \frac{{\Omega ^{p} }}
{{\omega ^p }}\quad \omega  \geqslant \Omega
\end{gathered}  \right.
   \end{equation}
where $0<p<1$.

In equation (\ref{g2}) the quantity $g^2(\Omega)$ represents the
on-resonant coupling strength between qubit and  one-dimensional
waveguide, satisfying the strong coupling condition $\Gamma\ll
g(\Omega)\ll \Omega$.

\begin{figure}
  \includegraphics[width=8.0 cm]{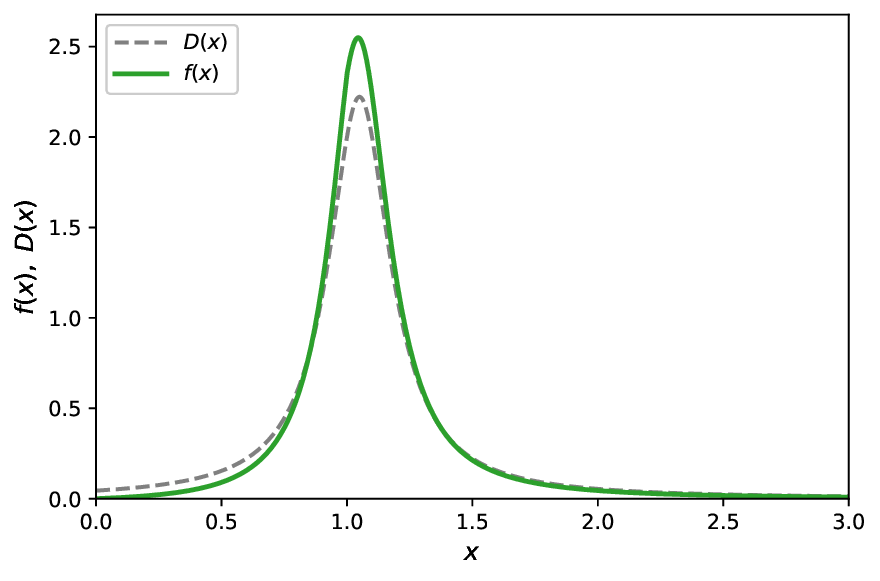}\\
  \caption{The comparison of the function $f(x)=f(\omega)/g^2(\Omega)$, solid line)
  with the Lorentzian $D(\omega)$ (dashed line). $x=\omega/\Omega$,
  $\Gamma/\Omega=0.15$, $\omega_0/\Omega=1.05$, $p=0.5$,
  normalization provides $\Lambda=g^2(\Omega)/0.39$, average frequency $\overline{\omega}/\Omega=1.17$.}\label{Fig1}
\end{figure}

The functions $f(\omega)$ and $D(\omega)$ are shown in
Fig.\ref{Fig1}  for $p=0.5$. We note that, in contrast to
$D(\omega)$, the function $f(\omega)$ vanishes at $\omega=0$.

\subsection{Calculation of the effective coupling strength $\Lambda$}

According to (\ref{Lam1}):
   \begin{equation}\label{Lam2}
\Lambda  = \int\limits_0^\infty  {} D(\omega)g^2(\omega)d\omega.
   \end{equation}
With the use of (\ref{ds1}) and (\ref{g2}) we obtain:
   \begin{equation}\label{Lam2b}
    \Lambda=g^2(\Omega)\gamma(I_1+I_2),
   \end{equation}
where:
   \begin{equation}\label{int}
\begin{gathered}
  I_1  = \int\limits_0^1 {dx} \frac{x}
{{(x - x_0 )^2  + \gamma ^2 }},
\\
  I_2  = \int\limits_1^\infty  {dx} \frac{1/x^p}
{\left( {(x - x_0 )^2  + \gamma ^2 } \right) },
\end{gathered}
   \end{equation}
and we introduced following parameters: $\gamma=\Gamma/\Omega$, $x=\omega/\Omega$, $x_0=\omega_0/\Omega$.

For $p=0.5$ these integrals can be expressed in the analytical
form:
   \begin{equation}\label{I1}
   \begin{gathered}
I_1  = \frac{1} {2}\ln \frac{{(1 - x_0 )^2  + \gamma ^2 }} {{x_0^2
+ \gamma ^2 }} \\+ \frac{{x_0 }} {\gamma }\left( {\arctan{(1 - x_0
)} + \arctan{(x_0)} } \right),
\end{gathered}
\end{equation}

\begin{equation}\label{I2}
I_2 = -\frac{1}{\gamma} \,\text{Im} \left[ \frac{1}{a}
\ln\frac{a+1}{a-1} \right],
   \end{equation}
where $a = \sqrt{x_0 + i\gamma}$.

\begin{figure}
  \includegraphics[width=8 cm]{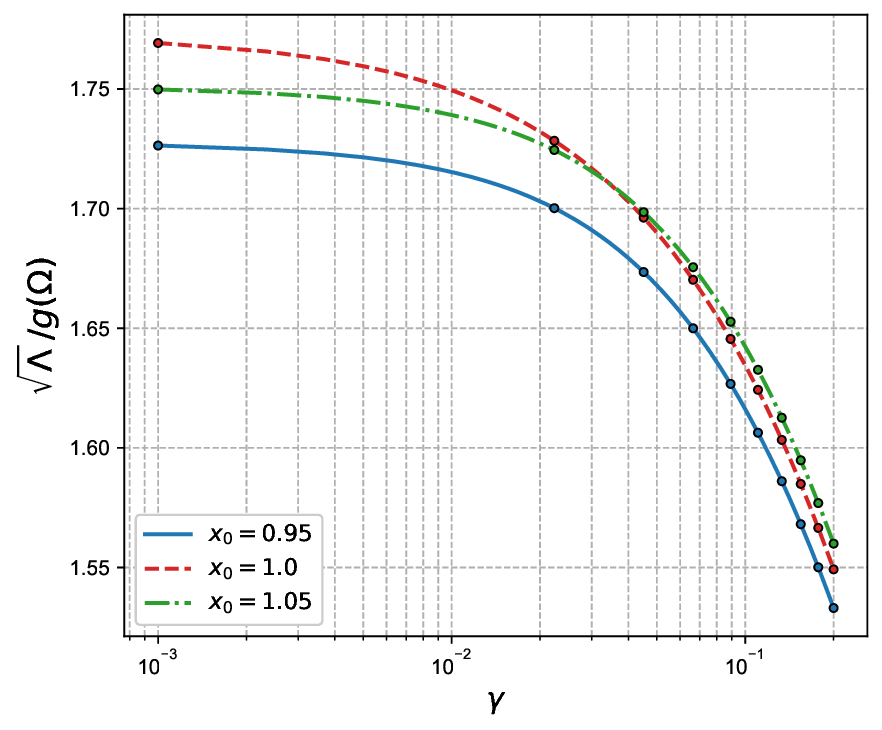}\\
  \caption{The dependence of the effective coupling
  $\sqrt{\Lambda}/g(\Omega)$ for $p=0.5$
  on the mode decay rate $\gamma=\Gamma/\Omega$ for different
  detunings $x_0=\omega_0/\Omega$. Solid (blue) line, $x_0=0.95$; dashed (red)
  line, $x_0=1$; dash-dotted (green) line, $x_0=1.05$
  The line values are numerically calculated from the integrals (\ref{int})
  while the markers on these lines are directly calculated from the analytical
   expressions (\ref{I1}), (\ref{I2}) .}\label{Fig2}
\end{figure}

\begin{figure}
  \includegraphics[width=8 cm]{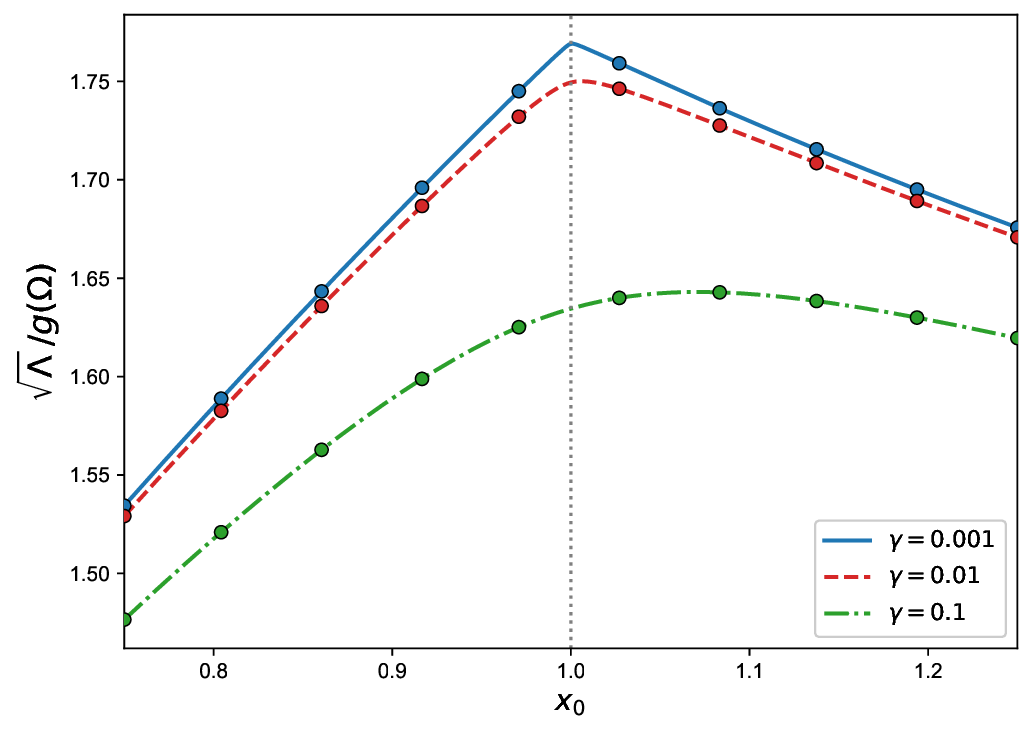}\\
  \caption{The dependence of the effective coupling
  $\sqrt{\Lambda}/g(\Omega)$ for $p=0.5$
  on the detuning $x_0=\omega_0/\Omega$  for different decay rates
  $\gamma=\Gamma/\Omega$. Solid (blue) line $\gamma=0.001$; dashed (red) line
   $\gamma=0.01$; dot-dashed (green) line $\gamma=0.1$. The line values are numerically calculated from
the integrals (\ref{int})
  while the markers on these lines are directly calculated from the analytical
   expressions (\ref{I1}), (\ref{I2}) .}\label{Fig3}
\end{figure}

Therefore, the effective coupling $\sqrt{\Lambda}$ depends on two
dimensionless  parameters, $x_0=\omega_0/\Omega$ and
$\gamma=\Gamma/\Omega$ as shown in Fig. \ref{Fig2} and Fig.
\ref{Fig3}.

It should be noted that in the parameter region of $x_0$ and
$\gamma$ shown in these figures, the effective coupling
$\sqrt{\Lambda}$ exceeds the single-mode coupling strength
$g(\Omega)$.

It is seen from Fig.\ref{Fig2} and Fig.\ref{Fig3} that the
   coupling strength increases as the decay rate of the mode
   $\gamma$ tends to zero. Physically, this is reasonable since the
   maximum coupling strength should be observed for a lossless
   waveguide. From analysis of integrals (\ref{int}) we
   find that for $p=0.5$ and $\gamma\rightarrow 0$ the coupling
   strength behaves as
    $\sqrt{\Lambda}/g(\Omega)\rightarrow \sqrt{x_0\pi}$ for $0<x_0\leq
    1$, and $\sqrt{\Lambda}/g(\Omega)\rightarrow \sqrt{\pi}x_0^{-1/4}$
    for $x_0>1$.

\subsection{Calculation of the average frequency $\overline{\omega}$}

From (\ref{avfr}) and (\ref{ds}) the average frequency can be
written as:
   \begin{equation}\label{avfr1}
\overline{\omega}  = \int\limits_0^\infty  {d\omega \,\omega
f(\omega )}.
   \end{equation}

Then, using the equations (\ref{ds}), (\ref{ds1}), (\ref{g2}), and
(\ref{Lam2b}), we can rewrite it further as:
   \begin{equation}\label{avfr2}
\frac{{\bar \omega }} {\Omega }=\frac{I_3+I_4}{I_1+I_2},
   \end{equation}
where:
   \begin{equation}\label{Int34}
\begin{gathered}
  I_3  = \int\limits_0^1 {dx} \frac{{x^2 }}
{{(x - x_0 )^2  + \gamma ^2 }},
\\
  I_4  = \int\limits_1^\infty  {dx} \frac{x^{1 - p}}
{{(x - x_0 )^2  + \gamma ^2 }}.
\end{gathered}
   \end{equation}

   \begin{table}
\caption{Numerically calculated coefficients of polynom (\ref{pol}) for three parameters of $p$.}
\label{table}
\begin{tabular}{c|c|c|c}
\toprule
 & $c_1$ & $c_2$ & $c_3$ \\
\hline
$p = 0.25$ & 1.472700 & 0.997056 & 0.675032 \\
$p = 0.5$ & 0.730007 & 0.583279 & 0.466043 \\
$p = 0.75$ & 0.438357 & 0.441358 & 0.444379 \\
\botrule
\end{tabular}
   \end{table}

By the numerical simulation of the integrals (\ref{Int34}) we show
that the dependence of the average frequency $\overline{\omega}$
on $\gamma=\Gamma/\Omega$ can be approximated by the polynomial,
the first few terms of which are given below:
   \begin{equation}\label{pol}
    \overline{x}=x_0+c_1\gamma+c_2\gamma^2+c_3\gamma^3
   \end{equation}
where $\overline{x}=\overline{\omega}/\Omega$,
$\overline{x_0}=\overline{\omega_0}/\Omega$. Calculated coefficients for three parameters $p = 0.25$, $p=0.5$ and $p=0.75$ can be found in Table. \ref{table}.

\begin{figure}
  \includegraphics[width=8.0 cm]{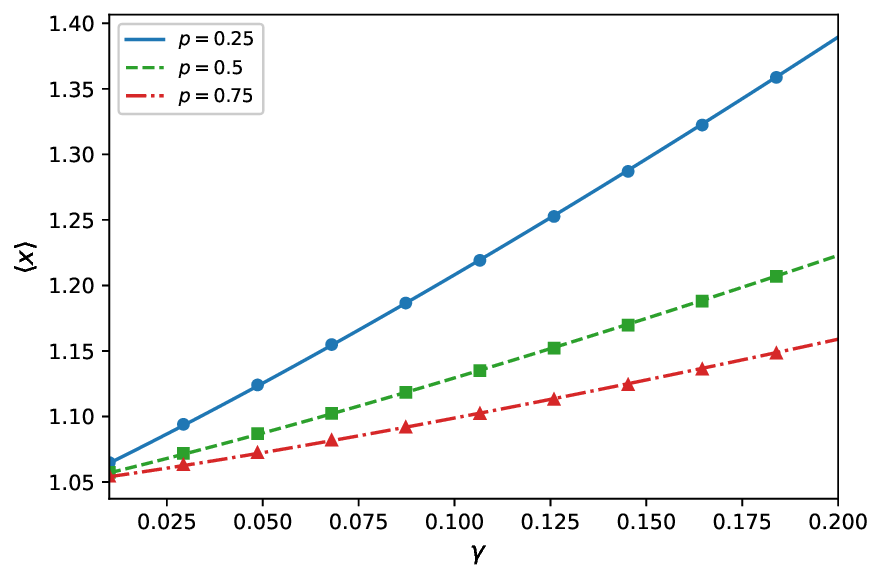}\\
  \caption{The dependence of the average
  frequency $\langle x\rangle=\overline{\omega}/\Omega$
  on the halfwidth $\gamma=\Gamma/\Omega$ of the function $f(\omega)$.
Solid (blue) line $p=0.25$; dashed (green) line
   $p=0.5$; dot-dashed (red) line $p=0.75$. The line values
   are numerically calculated from
 integrals in the expression (\ref{avfr2})
  while the markers on these lines are calculated from the
  polynomial expression (\ref{pol}).}\label{Fig4}
\end{figure}

For $\gamma<0.2$ the expression (\ref{pol}) provides a very good
approximation to the expression (\ref{avfr2}), as shown in
Fig.\ref{Fig4}.

As is seen in Fig.\ref{Fig4}, there exists the range of parameters,
$p>0.5, \gamma<0.1$, where the average frequency
$\overline{\omega}$ is close to the qubit frequency $\Omega$,
satisfying the condition $\overline{\omega}-\Omega \ll \Omega$.

\section{Multiphoton continuum Fock states}

Below we denote the full set of the states of the qubit-photon
system as $ \left| {s,\{n}\} \right\rangle $, where $|s\rangle$
stands for the ground, $|g\rangle$ or excited state $|e\rangle$ of
a qubit, and there are $n$ photons in a waveguide.

For  multiphoton continuum $n$ photons are distributed over all
frequencies, therefore, we may construct the multiphoton Fock
state as \cite{Mandel1995}:
   \begin{equation}\label{fock}
\left| {s,\{ n \}  } \right\rangle  = \left| s \right\rangle
\otimes \prod\limits_{\omega } {\left| {n (\omega )} \right\rangle
},
   \end{equation}
where $n(\omega)$ is the number of photons in the frequency mode
$\omega$:
   \begin{equation}\label{fock1}
\left| {n (\omega )} \right\rangle  = \frac{{\left( {a^\dagger
(\omega )} \right)^{n (\omega )} }} {{\sqrt {n (\omega )!}
}}\left| {0} \right\rangle ,
   \end{equation}
where $|0\rangle$ is the multimode vacuum state with no photons in
all modes. The total number of photons is obtained by
integration over all frequencies $ n  = \int\limits_0^\infty
{d\omega D(\omega )n (\omega )} $  with $D$ representing the
density of modes per unit energy. It is worth noting that the
state (\ref{fock}) for fixed $n$ represents the infinite ensemble
of Fock states with $n$ being distributed over all frequencies in
different combinations.

The states (\ref{fock}) satisfy the completeness relation:
  \begin{equation}\label{2.0}
\sum\limits_{s,\{n\} } {} \left| {s,\{n\}  } \right\rangle
\left\langle {s,\{n\}  } \right| = 1,
   \end{equation}
and the normalizing condition:
    \begin{equation}\label{fock2}
\left\langle {s,\{ n \}  |s',\{ m \}  } \right\rangle = \delta
_{s,s'} \prod\limits_{\omega } {} \delta _{n (\omega ),m (\omega
)} .
   \end{equation}

For example, for multimode case the state $  \left| {s,\{1\} }
\right\rangle $, where $s=g,e$ consists of infinite number of
states each of which has a single photon in $k$th mode:
\begin{equation}\label{s}
     \left| s,\{1\} \right\rangle   = \left\{ \begin{gathered}
  \left| {s,1_{k_1 } } \right\rangle  \equiv \left| {s,1_{k_1 } ,0,0,0......0} \right\rangle , \hfill \\
  \left| {s,1_{k_2 } } \right\rangle  \equiv \left| {s,0,1_{k_2 } ,0,0,0.....0} \right\rangle , \hfill \\
  \left| {s,1_{k3} } \right\rangle  \equiv \left| {s,0,0,1_{k_3 } ,0,0,0......0} \right\rangle , \hfill \\
  ................................... \hfill \\
\end{gathered}  \right.
   \end{equation}

\section{Multiphoton continuum Rabi oscillations}

We define the following multiphoton $A$-number states:
\begin{equation}\label{32}
\left| n \right\rangle_A  = \frac{1} {{\sqrt {n!} }}(A^\dagger  )^n
\left| 0 \right\rangle
\end{equation}
where $A^\dagger$ and $A$ are defined in (\ref{4}).

Due to commutator (\ref{Com}) and the vacuum condition
$A|0\rangle=0$ the usual boson relations hold:
\begin{equation}\label{33}
\begin{gathered}
  A\left| n \right\rangle_A  = \sqrt n \left| {n - 1} \right\rangle_A,
\\
  A^\dagger \left| n \right\rangle_A  = \sqrt {n + 1} \left| {n + 1} \right\rangle_A,
\\
_A\langle n|m\rangle_A=\delta_{n,m},
\end{gathered}
\end{equation}
where $n$ is the total number of photons distributed over all
frequencies in the state $|n\rangle_A$.

It should be pointed out that, strictly speaking, the $A$- number
states (\ref{32}) are not the Fock number states (\ref{fock}). The
action of the photon number operator $a^\dagger(\omega)a(\omega)$ on
multimode Fock state (\ref{fock}) provides the number of photons
$n(\omega)$ in the frequency mode $\omega$. However, for
$A$-number states (\ref{32}):
   \begin{equation}\label{f}
\begin{gathered}
  a(\omega )\left| n \right\rangle_A  = \frac{{g(\omega )}}
{{\sqrt \Lambda  }}\sqrt n \left| {n - 1} \right\rangle_A ,
\\
  a^\dagger  (\omega )\left| n \right\rangle_A  = \frac{{g(\omega )}}
{{\sqrt \Lambda  }}\sqrt {n + 1} \left| {n + 1} \right\rangle_A .
\end{gathered}
   \end{equation}

Therefore, for $A$-number states:
   \begin{equation}\label{ns}
a^\dagger  (\omega )a(\omega )\left| n \right\rangle_A  = \frac{{g^2
(\omega )}} {\Lambda }n\left| n \right\rangle_A,
   \end{equation}
where $g^2(\omega)n/\Lambda$ is the fraction of the photons with
the frequency mode $\omega$ in the number state $|n\rangle_A$. Due
to (\ref{Lam}), the total number of photons in the state
$|n\rangle_A$ is equal to $n$.

In this section, we examine the exchange of a single photon
between two composite states: the excited qubit with \(n-1\)
photons in the state  \(|n-1\rangle_A\), and the ground-state
qubit with \(n\) photons in the state \(|n\rangle_A\).

The trial wavefunction is taken in the form
   \begin{equation}\label{34}
\Psi _n  = A_n \left| {e,n - 1} \right\rangle_A  + B_n \left|
{g,n} \right\rangle_A.
   \end{equation}
The function (\ref{34}) is the eigenfunction of the operator of
number of excitations: $\widehat{N}\Psi_n=n\Psi_n$. The eigenvalue
$q_n$ of the operator $\widehat{Q}$ is obtained from the
Schr\"odinger equation $\widehat{Q}\Psi_n=q_n\Psi_n$, which results
in two equations for $A_n$ and $B_n$:
   \begin{equation}\label{35}
\begin{gathered}
  A_n (n - 1)G + B_n \sqrt {\Lambda n}  = q_n A_n,
\\
  A_n \sqrt {\Lambda n}  + B_n nG = q_n B_n.
\end{gathered}
\end{equation}

From these equations and orthonormal properties of  the function
(\ref{34}) we find:
   \begin{equation}\label{36}
q_{n \pm }  = \frac{{2n - 1}} {2}G \pm \frac{1} {2}\sqrt {G^2  +
4\Lambda n},
   \end{equation}
   \begin{equation}\label{37}
\begin{gathered}
  \Psi _{n + }  = \cos \theta _n \left| {e,n - 1} \right\rangle_A  + \sin \theta _n \left| {g,n} \right\rangle_A ,
  \\
  \Psi _{n - }  = \sin \theta _n \left| {e,n - 1} \right\rangle_A  - \cos \theta _n \left| {g,n} \right\rangle_A,
\end{gathered}
   \end{equation}
where:
   \begin{equation}\label{38}
\tan\theta _n  = \frac{{q_{n + }  - (n - 1)G}} {{\sqrt {\Lambda n}
}}.
   \end{equation}

Now we construct the time-dependent wave function with initially
excited qubit and $n-1$ photons: $\Psi(0)=|e,n-1\rangle_A$:
   \begin{equation}\label{39}
\begin{gathered}
\Psi (t) = \left( {e^{ - iE_{n+}  t} \cos ^2 \theta _n  + e^{ - iE_{n - } t} \sin^2 \theta _n } \right)\left| {e,n - 1} \right\rangle_A
\\
+ \cos \theta _n \sin \theta _n \left( {e^{ - iE_{n + } t}  - e^{ - iE_{n - } t} } \right)\left| {g,n} \right\rangle_A,
\end{gathered}
   \end{equation}
where $ E_{n \pm }  = q_{n \pm }  + n\Omega$.

The probabilities $P_e(t)$, $P_g(t)$ and $W(t)$ are similar to
those obtained in (\ref{27}), (\ref{28}), (\ref{30}):
   \begin{equation}\label{40}
\begin{gathered}
P_e (t) = \left| {e^{ - iE_{n + } t} \cos ^2 \theta_n  +
e^{ - iE_{n - } t} \sin ^2 \theta_n } \right|^2
\\
= 1 - 4\sin ^2 \theta_n \cos ^2 \theta_n \sin ^2 \frac{{E_{n + } -
E_{n - } }}
{2}t,
\end{gathered}
   \end{equation}
   \begin{equation}\label{41}
\begin{gathered}
  P_g (t) = \cos ^2 \theta_n \sin ^2 \theta_n \left|
  {\left( {e^{ - iE_{n + } t}  - e^{ - iE_{n - } t} } \right)} \right|^2
\\
   = 4\sin ^2 \theta_n \cos ^2 \theta_n \sin ^2 \frac{{E_{n + }  - E_{n - } }}
{2}t,
\end{gathered}
   \end{equation}
   \begin{equation}\label{42}
\begin{gathered}
W(t) = P_e (t) - P_g (t)
\\
= 1 - 8\cos ^2 \theta_n \sin ^2
\theta_n \sin ^2 \frac{{E_ {n+}   - E_{n -}  }} {2}t,
\end{gathered}
   \end{equation}
where:
\begin{equation}\label{43}
E_{n + }  - E_{n - }  = q_{n + }  - q_{n - }  = \sqrt {G^2  +
4\Lambda n}.
\end{equation}

Therefore, we have demonstrated that in our case the system's
dynamics can be reduced to an effective two-level problem
involving the qubit and multiphoton collective excitations.

\section{Multiphoton Rabi oscillations for initial continuum coherent state}

In this section we investigate the evolution of the wave function
when the qubit is initially in the excited state and the initial
field is a continuum coherent state \cite{Blow1990}. In this case,
the time dependence of the wave-function amplitude exhibits an
oscillatory process involving the exchange of a photon between
multphoton continuum states differing by one photon.

\subsection{Continuum field coherent states}

We define the continuum mode creation, $a^\dagger_{\alpha}$ and
destruction, $a_{\alpha}$ operators as follows:
   \begin{equation}\label{45}
\begin{gathered}
  a_\alpha   = \frac{1}
{{\sqrt {\left\langle {n_\alpha  } \right\rangle } }}\int\limits_0^\infty  {d\omega } \alpha ^* (\omega )a(\omega ),
\\
  a_\alpha ^\dagger   = \frac{1}
{{\sqrt {\left\langle {n_\alpha  } \right\rangle } }}\int\limits_0^\infty  {d\omega } \alpha (\omega )a^\dagger  (\omega ),
\end{gathered}
   \end{equation}
where:
   \begin{equation}\label{46}
\left\langle {n_\alpha  } \right\rangle  = \int\limits_0^\infty
{d\omega } \left| {\alpha (\omega )} \right|^2,
   \end{equation}
and their commutation relation is:
   \begin{equation}\label{47}
    [a_{\alpha},a^\dagger_{\alpha}]=1.
   \end{equation}

The quantity $\alpha(\omega)$ is arbitrary quadratically
integrable function of the frequency. The continuum mode coherent
state for the function $\alpha(\omega)$ can be written as follows
\cite{Blow1990}:
   \begin{equation}\label{44}
\left| {\alpha (\omega )} \right\rangle  = \exp \left(
{\left\langle \sqrt{{n_\alpha  }} \right\rangle (a_\alpha ^\dagger   -
a_\alpha  )} \right),
   \end{equation}
with the normalizing condition:
   \begin{equation}\label{norm_coh}
    \langle\alpha(\omega)|\alpha(\omega)\rangle=1.
  \end{equation}

Applying to (\ref{44}) the Campbell formula $ e^{A + B}  = e^A e^B
e^{ - \frac{1} {2}[A,B]}$, which is valid if the commutator $[A,B]$
is a $c$-number, we can express the coherent state (\ref{44}) in the
following form:
   \begin{equation}\label{48}
\begin{gathered}
  \left| {\alpha (\omega )} \right\rangle  = e^{ - \frac{1}
{2}\left\langle \sqrt{{n_\alpha  }} \right\rangle } e^{\left\langle \sqrt{{n_\alpha}  } \right\rangle a_\alpha ^\dagger  } \left| 0 \right\rangle
\\
   = e^{ - \frac{1}
{2}\left\langle {n_\alpha  } \right\rangle } \sum\limits_{n =
0}^\infty  {} \frac{{\left( {\sqrt {\left\langle {n_\alpha  }
\right\rangle } } \right)^n \left( {a_\alpha ^\dagger  } \right)^n }}
{{n!}}\left| 0 \right\rangle.
\end{gathered}
   \end{equation}

It is not difficult to show that the coherent state (\ref{48}) is
the eigenstate of both the continuum mode destruction operators
$\alpha(\omega)$ and $a_{\alpha}$:
   \begin{equation}\label{51}
    a(\omega')|\alpha(\omega)\rangle=\alpha(\omega')|\alpha(\omega)\rangle,
   \end{equation}
   \begin{equation}\label{52a}
a_\alpha  \left| {\alpha (\omega )} \right\rangle  = \sqrt
{\left\langle {n_\alpha  } \right\rangle } \left| {\alpha (\omega
)} \right\rangle.
   \end{equation}

Therefore, $[a(\omega),a_{\alpha}=0]$ and:
   \begin{equation}\label{alpha}
a(\omega)=\frac{\alpha(\omega)}{\sqrt{\langle
n_{\alpha}\rangle}}a_{\alpha}.
   \end{equation}

It follows also from (\ref{52a}) that $\left\langle {\alpha
(\omega )} \right|a_\alpha ^\dagger  a_\alpha  \left| {\alpha (\omega
)} \right\rangle  = \left\langle {n_\alpha  } \right\rangle$ where
$\langle n_{\alpha}\rangle$ is the average number of photons in
the coherent state given in (\ref{46}).

Next, we define the $a$- number state basis for coherent state
$|\alpha(\omega)\rangle$:
   \begin{equation}\label{49}
\left| {n } \right\rangle_a   = \frac{1} {{\sqrt {n!} }}\left(
{a_\alpha ^\dagger  } \right)^n \left| 0 \right\rangle,
   \end{equation}
with the properties:
   \begin{equation}\label{53}
\begin{gathered}
  a_\alpha  \left| {n  } \right\rangle_a  = \sqrt n \left| {n - 1  } \right\rangle_a,
\\
  a_\alpha ^\dagger  \left| {n  } \right\rangle_a  = \sqrt {n + 1} \left| {n + 1  } \right\rangle_a ,
\\
 _a \left\langle {n  |m } \right\rangle _a  = \delta _{n,m}.
\end{gathered}
   \end{equation}

From (\ref{alpha}) we also obtain:
   \begin{equation}\label{nal}
\begin{gathered}
  a(\omega )\left| {n  } \right\rangle_a  = \frac{{\alpha (\omega )}}
{{\sqrt {\left\langle {n_\alpha  } \right\rangle } }}\sqrt n \left| {n - 1  } \right\rangle_a,
\\
  a^\dagger  (\omega )\left| {n  } \right\rangle_\alpha  = \frac{{\alpha (\omega )}}
{{\sqrt {\left\langle {n_\alpha  } \right\rangle } }}\sqrt {n + 1} \left| {n + 1  } \right\rangle_a,
\end{gathered}
   \end{equation}
   \begin{equation}\label{ff}
a^\dagger  (\omega )a(\omega )\left| {n  } \right\rangle_a  =
\frac{{\left| {\alpha (\omega )} \right|^2 }} {{\left\langle
{n_\alpha  } \right\rangle }}n\left| {n  } \right\rangle_a.
   \end{equation}

In terms of $a$- number states $|n\rangle_a$ the coherent state
(\ref{48}) can be written as:
   \begin{equation}\label{50}
\left| {\alpha (\omega )} \right\rangle  = e^{ - \frac{1}
{2}\left\langle {n_\alpha  } \right\rangle } \sum\limits_{n =
0}^\infty  {} \frac{{\left( {\sqrt {\left\langle {n_\alpha  }
\right\rangle } } \right)^n }} {{\sqrt {n!} }}\left| {n  }
\right\rangle_a.
   \end{equation}

\subsection{Rabi oscillations}

In this subsection we find the time-dependent wave function
$\Psi(t)$ for which initially the qubit is excited and the field
is in a coherent state: $\Psi(0)=|e\rangle\bigotimes
|\alpha(\omega)\rangle\equiv |e,\alpha(\omega)\rangle$.

We construct the trial wavefuncton as a
superposition of $a$-number states (\ref{49}):
   \begin{equation}\label{51b}
\Psi _{n}  = C_n \left| {e,n - 1  } \right\rangle_a + D_n \left|
{g,n  } \right\rangle_a.
   \end{equation}
It can be shown that the states $| {e,n - 1  } \rangle_a $ and $|
{g,n  } \rangle_\alpha$ are eigenstates for the operator
$\widehat{N}$ with the eigenvalue $n$. Therefore,
$\widehat{N}\Psi_{n}=n\Psi_{n}$. The eigenvalue $q_{n_{\alpha}}$
of the operator $\widehat{Q}$ can be found from the Schr\"odinger
equation $\widehat{Q}\Psi_{n}=q_{n_{\alpha}}\Psi_{n}$ which
provides two equations for $C_n$ and $D_n$ (see Appendix B):
   \begin{equation}\label{52}
C_n \left( {(n - 1)\Delta  - q_{n_\alpha  } } \right) + D_n
\frac{F} {{\sqrt {\left\langle {n_\alpha  } \right\rangle }
}}\sqrt n  = 0,
   \end{equation}
   \begin{equation}\label{53b}
C_n \frac{{F^* }} {{\sqrt {\left\langle {n_\alpha  } \right\rangle
} }}\sqrt n  + D_n (n\Delta  - q_{n_\alpha  } ) = 0,
   \end{equation}
where:
   \begin{equation}\label{54}
\Delta  = \frac{1} {{\left\langle {n_\alpha  } \right\rangle
}}\int\limits_0^\infty  {d\omega } (\omega  - \Omega )\left|
{\alpha (\omega )} \right|^2,
   \end{equation}
   \begin{equation}\label{F}
F = \int\limits_0^\infty  {d\omega } {d(\omega) }\alpha (\omega
)g(\omega ),
   \end{equation}
and $d(\omega)$ is the photon density of states.

From equations (\ref{52}), (\ref{53b}) together with the
normalizing condition $|C_n|^2+|D_n|^2=1$ we obtain:
   \begin{equation}\label{55}
q_{n_a  \pm }  = \frac{{2n - 1}} {2}\Delta  \pm \frac{1}{2}\sqrt
{\Delta ^2 + \frac{{4\left| F \right|^2 n}} {{\left\langle
{n_\alpha  } \right\rangle }}},
   \end{equation}
   \begin{equation}\label{56}
\begin{gathered}
  \Psi _{n_\alpha   + }  = A_{n + } \left| {e,n - 1  }
  \right\rangle_a
   + B_{n + } \left| {g,n  } \right\rangle_a,
\\
  \Psi _{n_\alpha   - }  = A_{n - } \left| {e,n - 1  }
  \right\rangle_a
   + B_{n - } \left| {g,n  } \right\rangle_a.
\end{gathered}
   \end{equation}

The coefficients $A_{n\pm }$, $B_{n\pm }$ can be found from the
orthonormality conditions:
   \begin{equation}\label{57}
\begin{gathered}
  A_{n + }  = \cos \theta _{n_\alpha  } ,\;B_{n + }  = \sin \theta _{n_\alpha  },
\\
  A_{n - }  = \sin \theta _{n_\alpha  } ,\;B_{n - }  =  - \cos \theta _{n_\alpha  },
\end{gathered}
   \end{equation}
where:
   \begin{equation}\label{58}
\tan\theta _{n_\alpha  }  = \frac{{q_{n_\alpha   + }  - (n -
1)\Delta }} {{F\sqrt n }}\sqrt {\left\langle {n_\alpha  }
\right\rangle }.
   \end{equation}

Time-dependent wavefunction with the initial condition $\Psi_n (0)
= \left| {e,n - 1 } \right\rangle_a$ can be constructed similar to
expression (\ref{39}):
   \begin{equation}\label{59}
\begin{gathered}
  \Psi _{n_\alpha} (t) = \left( {e^{ - iE_{n_\alpha + }   t} \cos^2 \theta _{n_a }
    + e^{ - iE_{n_\alpha   - } t} \sin^2 \theta _{n_\alpha  } } \right)\left| {e,n - 1  } \right\rangle_a
\\
   + \cos \theta _{n_\alpha  }
   \sin \theta _{n_\alpha  } \left( {e^{ - iE_{n_\alpha   + } t}  - e^{ - iE_{n_\alpha   - } t} } \right)\left| {g,n  } \right\rangle_a,
\end{gathered}
   \end{equation}
where $E_{n_\alpha   \pm }  = q_{n_\alpha   \pm }  + n\Omega$. Now
it is not difficult to find wavefunction with the initial
condition $ \Psi_{n_{\alpha}} (0) = |e\rangle\bigotimes |\alpha
(\omega)\rangle$, where $|\alpha (\omega )\rangle$ is given in
(\ref{50}):
   \begin{equation}\label{60}
\Psi (t) = ^{ - \frac{1} {2}\left\langle {n_\alpha  }
\right\rangle } \sum\limits_{n = 1}^\infty  {} \frac{{\left(
{\sqrt {\left\langle {n_\alpha  } \right\rangle } } \right)^{n -
1} }} {{\sqrt {(n - 1)!} }}\Psi _{n_{\alpha}} (t).
   \end{equation}

If $ \frac{{\left| F \right|\sqrt n }} {{\sqrt {\left\langle
{n_\alpha  } \right\rangle } }} >  > \Delta $ then $\tan \theta
_{n_\alpha  }  = 1 $ and we obtain for (\ref{60}):
   \begin{equation}\label{61}
\begin{gathered}
  \Psi (t) = ^{ - \frac{1}
{2}\left\langle {n_\alpha  } \right\rangle } \sum\limits_{n =
1}^\infty  {} \frac{{\left( {\sqrt {\left\langle {n_\alpha  }
\right\rangle } } \right)^{n - 1} }}
{{\sqrt {(n - 1)!} }}e^{ - in\Omega t} \times
\\
\left( \begin{gathered}
  \cos \left( {\frac{{\left| F \right|\sqrt n }}
{{\sqrt {\left\langle {n_\alpha  } \right\rangle } }}t}
\right)\left| {e,n - 1  } \right\rangle_a  - i\sin \left(
{\frac{{\left| F \right|\sqrt n }}
{{\sqrt {\left\langle {n_\alpha  } \right\rangle } }}t} \right)\left| {g,n  } \right\rangle_a
\end{gathered}  \right).
\end{gathered}
   \end{equation}

From (\ref{61}) we obtain the inversion
$W(t)=\langle\Psi(t)|\sigma_z|\Psi(t)\rangle$:
   \begin{equation}\label{62}
W(t) =e ^{ - \left\langle {n_\alpha  } \right\rangle }
\sum\limits_{n = 1}^\infty  {} \frac{{\left( {\left\langle
{n_\alpha  } \right\rangle } \right)^{n - 1} }} {{(n - 1)!}}\cos
\left( {\frac{{2\left| F \right|\sqrt n }} {{\sqrt {\left\langle
{n_\alpha  } \right\rangle } }}t} \right).
   \end{equation}

It should be noted that, in fact, the Rabi frequency
\(2|F|\sqrt{n}/\langle n_{\alpha}\rangle\) in (\ref{62}) does not
depend on the magnitude of \(\langle n_{\alpha}\rangle\). This
follows from the fact that the function \(\alpha(\omega)\), which
generates the coherent state \(|\alpha(\omega)\rangle\), is
normalized to \(\sqrt{\langle n_{\alpha}\rangle}\), as follows
from (\ref{46}). That is, we can write
\(\alpha(\omega)=\sqrt{\langle
n_{\alpha}\rangle}\,\tilde{\alpha}(\omega)\), where
\(\tilde{\alpha}(\omega)\) is independent of \(\langle
n_{\alpha}\rangle\) with the normalization:
   \begin{equation}\label{norm1}
\int\limits_0^\infty {d\omega \left| {\tilde \alpha (\omega )}
\right|^2 }  = 1.
   \end{equation}
Therefore, the ratio \(\alpha(\omega)/\sqrt{\langle
n_{\alpha}\rangle}\) does not depend on the magnitude of \(\langle
n_{\alpha}\rangle\) and the Rabi frequency can be written in the
following form \cite{Dzs2016}:
   \begin{equation}\label{Rabi_coh}
\Omega _R  = 2\sqrt n \int\limits_0^\infty  {d\omega }\, d(\omega)
\widetilde{g}(\omega )\tilde \alpha (\omega ).
   \end{equation}
where
\begin{equation}\label{B12}
    \widetilde{g}(\omega)=\sqrt{\frac{\omega d}{4\pi\varepsilon_0\hbar A}}
\end{equation}
 and $d(\omega)$ is defined as follows:

\begin{equation}\label{ds4}
    d(\omega)=\frac{\sqrt{\Gamma}}{\omega-\omega_0+i\Gamma}
\end{equation}

As is seen $d(\omega)$ is closely related to the photonic density
of states (\ref{ds1}), $D(\omega)=|d(\omega)|^2$.

The integral in (\ref{Rabi_coh}) is the result of the
transformation to the continuum of the discrete sum $
\sum\limits_k {} g_k a_k^\dag$ which appears in the interaction
Hamiltonian (\ref{HJC}). This transformation is explained in
Appendix B.

 It should also be noted that, in contrast to the single-mode
case, here the Rabi frequency depends significantly on the shape
of the generator line \(\alpha(\omega)\) and on the decay rate of
the density of states of the photonic mode. In the following
subsection, we will calculate these dependencies for a Gaussian
form of the function \(\alpha(\omega)\).

\subsection{Dependence of Rabi frequency on the photon loss rate }

Here we calculate the dependence of the Rabi frequency on the loss
rate of the photonic mode for a Gaussian shape of the function
\(\widetilde{\alpha}(\omega)\), assuming that the coherent field
is generated by a laser pulse:
   \begin{equation}\label{Gauss}
 \tilde \alpha (\omega ) = \left( {\frac{2} {{\pi \sigma ^2 }}}
\right)^{1/4} \exp \left( { - \frac{{(\omega  - \omega _\alpha )^2
}} {{\sigma ^2 }}} \right),
   \end{equation}
where $\omega_a$ is the laser frequency, $\sigma$ is the width of
the spectral line.

The photonic density of states, $d(\omega)$ is taken in the form
(\ref{ds1}). The coupling strength is taken as follows (see
(\ref{g2})):
   \begin{equation}\label{gg}
\widetilde{g}(\omega ) = \widetilde{g}(\Omega )\left\{
\begin{gathered}
  \left( {\frac{\omega }
{\Omega }} \right)^{1/2} \quad \quad \omega  \leqslant \Omega
\\
  \left( {\frac{\Omega }
{\omega }} \right)^{p/2} \quad \omega  \geqslant \Omega
\end{gathered}  \right.
\end{equation}

\begin{figure}
  \includegraphics[width=8 cm]{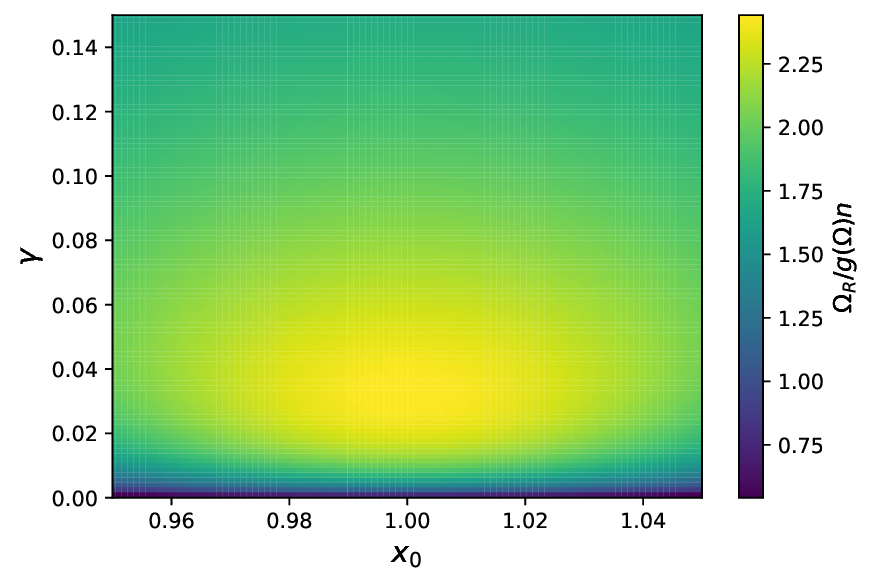}\\
  \caption{Color map of the dependence of the Rabi frequency $\Omega_R/\widetilde{g}(\Omega)n$
  on the photon loss rate $\gamma=\Gamma/\Omega$ and
  the detuning $x_0=\omega_0/\Omega$. $\omega_a=\Omega$. }\label{Fig5}
\end{figure}

The dependence of the dimensionless quantity $\Omega_R/g(\Omega)n$
on the the photon loss rate $\gamma=\Gamma/\Omega$ and
the detuning $x_0=\omega_0/\Omega$ for $\omega_a=\Omega$ is shown
in Fig.\ref{Fig5}.

These numerical results for a Gaussian lineshape of coherent state
showed a strong dependence of the Rabi frequency on the loss rate
of the photonic mode, confirming that the dissipation inherent in
open waveguides significantly influences the coherent dynamics.

\section{Conclusion}

In this work, we have developed a comprehensive theoretical
framework for describing quantum Rabi oscillations of a qubit
strongly coupled to the continuous spectrum of a one-dimensional
open waveguide. By employing the multimode Jaynes-Cummings
Hamiltonian within the rotating-wave approximation and utilizing
collective operators, we have demonstrated that the system's
dynamics can be reduced to an effective two-level problem
involving the qubit and multiphoton collective excitations.

Our analysis of the single-photon case revealed that the qubit
does not exchange energy with a single discrete mode, but rather
with a superposition of continuum modes. The resulting Rabi
frequency is determined by the effective coupling strength
\(\sqrt{\Lambda}\), which is defined as an integral over the
spectral density of the waveguide. We calculated this effective
coupling for a realistic Lorentzian density of states of photon
modes and a frequency-dependent coupling function, showing that
\(\sqrt{\Lambda}\) depends crucially on the mode loss rate and on
the detuning between the qubit frequency and the fundamental mode
frequency. We also derived an analytical expression for the
weighted average frequency and found that, for certain parameters,
it closely matches the qubit frequency.

For multiphoton states, we introduced the formalism of
\(A\)-number states, in which photons are distributed over
frequency modes according to the qubit-waveguide coupling. In the
limit of strong coupling, the Rabi frequency for the \(n\)-photon
manifold scales as \(\sqrt{\Lambda n}\), analogous to the
single-mode case. Furthermore, we investigated the dynamics for an
initial continuum coherent state, which allowed us to describe the
periodic exchange of photons between coherent states differing by
one photon. Crucially, we found that the Rabi frequency in this
case is not solely a function of the coupling strength, but
depends strongly on the specific shape of the coherent-state
spectral function and on the photon density of states. Our
numerical results for a Gaussian coherent state also showed a
strong dependence of the Rabi frequency on the loss rate of the
photonic mode, confirming that the dissipation inherent in open
waveguides fundamentally modifies the coherent dynamics.

Our findings provide a rigorous analytical foundation for
understanding strong-coupling phenomena in waveguide quantum
electrodynamics and highlight the distinct features of Rabi
oscillations in open systems compared to traditional cavity QED.

\begin{acknowledgments}

Y.S.G. acknowledges fruitful discussions with A. N. Sultanov. The
work is supported by the Ministry of Science and Higher Education
of Russian Federation under the project FSUN-2026-0004.

\end{acknowledgments}

\appendix

\section{Derivation of equation (\ref{operU})}\label{A}

Direct calculations show that:
   \begin{equation}\label{C2}
H_{_{JC} }^{2n + 1} \left| {e,0 } \right\rangle  =
\sqrt{\Lambda}\Lambda^n  A^\dagger \left|{{g,0 }} \right\rangle ,\quad n
= 0,1,2,3.....
   \end{equation}
   \begin{equation}\label{C3}
H_{_{JC} }^{2n} \left| {e,0 } \right\rangle  = \Lambda^n \left|
{e,0 } \right\rangle ,\quad n = 1,2,3.....
   \end{equation}

From the series expansion of the exponent we obtain:
   \begin{equation}\label{C6}
\begin{gathered}
  e^{-iH_{JC} t } \left| {e,0 } \right\rangle  =
  \sum\limits_{n = 0}^\infty  {} \frac{{\left( {-iH_{JC} t } \right)^n }}
{{n{\text{!}}}}\left| {e,0 } \right\rangle
\\
   = \sum\limits_{n = 0}^\infty  {} \frac{{\left( {-it } \right)^{2n} }}
{{\left( {2n} \right){\text{!}}}}\Lambda ^n \left| {e,0 }
\right\rangle  + \sum\limits_{n = 0}^\infty  {} \frac{{\left( {-it
} \right)^{2n + 1} }} {{\left( {2n + 1}
\right){\text{!}}}}\sqrt{\Lambda}\Lambda ^n | {g,0 } \rangle .
\end{gathered}
   \end{equation}
The expression (\ref{C6}) can be rewritten as:
   \begin{equation}\label{C13}
\begin{gathered}
  e^{ - iH_{JC} t } \left| {e,0} \right\rangle
\\
   = \cos \left( {\sqrt \Lambda  t } \right)\left| {e,0} \right\rangle
    - i\sin ( {\sqrt {\Lambda}  t })
 A^\dagger |{g,0} \rangle
\end{gathered}
   \end{equation}
which is the equation (\ref{operU}) from the main text.

\section{Transition to the continuum.
Comment on the expression (\ref{Rabi_coh})}\label{B}

The transition from summation over discrete variables to
integration over continuous variables in the one-dimensional case
for a flat spectrum is carried out according to the rule
\begin{equation}\label{B3}
\sum\limits_k {}  \to \frac{L} {{2\pi v_g }}\int\limits_0^{ +
\infty } {d\omega }
\end{equation}

where $L$ is one dimensional volume, $v_g$ is the group velocity
of electromagnetic waves, $L/2\pi v_g$ is the density of photon
states in the frequency domain.

In this case, the photon creation and annihilation operators are
transformed according to the rule \cite{Blow1990}
\begin{equation}\label{B4}
\begin{gathered}
  a_k  \to \sqrt {\frac{{2\pi v_g }}
{L}} a(\omega ) \hfill \\
  a_k^\dag   \to \sqrt {\frac{{2\pi v_g }}
{L}} a^\dag  (\omega ) \hfill \\
\end{gathered}
\end{equation}

which ensures that the commutation relations
\begin{equation}\label{B5}
  [a_k ,a_{k'}^\dag  ] = \delta _{k,k'}
\end{equation}
\begin{equation}\label{B6}
    a(\omega ),a^\dag  (\omega ')] = \delta (\omega  - \omega ')
\end{equation}

are satisfied. Thus
\begin{equation}\label{B7}
\begin{gathered}
  \sum\limits_k {} \omega _k a_k^\dag  a_k  \to \frac{L}
{{2\pi v_g }}\int\limits_0^\infty  {d\omega \omega } a_k^\dag  a_k  \hfill \\
   = \int\limits_0^\infty  {d\omega \omega } \left( {\sqrt {\frac{L}
{{2\pi v_g }}}\, a_k^\dag  } \right)\left( {\sqrt {\frac{L}
{{2\pi v_g }}}\, a_k } \right) = \int\limits_0^\infty  {d\omega \omega } a_{}^\dag  (\omega )a(\omega ) \hfill \\
\end{gathered}
\end{equation}
We see that  the density of states is absorbed by continuum mode
operators $a^\dag(\omega), a(\omega)$ in right hand side of
(\ref{B7}).

Consider now the transformation of the discrete sum  $
\sum\limits_k {} g_k a_k^\dag$ which appears in the interaction
Hamiltonian (\ref{HJC}). Here the coupling strength

 \begin{equation}\label{B8}
    g_k=\sqrt{\frac{\omega_k d}{2\varepsilon_0\hbar LA}}
\end{equation}
where $d$ is the off-diagonal matrix element of an atom dipole
operator, $A$ is the effective transverse cross section of
electromagnetic modes \cite{Dom2002}.

Using the rules (\ref{B2}) and (\ref{B3}) we obtain:

\begin{equation}\label{B9}
\begin{gathered}
  \sum\limits_k {} g_k a_k^\dag   \to \frac{L}
{{2\pi v_g }}\int\limits_0^\infty  {d\omega g_k } a_k^\dag   \hfill \\
   = \int\limits_0^\infty  {d\omega } \left( {\sqrt {\frac{L}
{{2\pi v_g }}} \,g_k } \right)\left( {\sqrt {\frac{L}
{{2\pi v_g }}} \,a_k^\dag  } \right) = \int\limits_0^\infty  {d\omega } g(\omega )a_{}^\dag  (\omega ) \hfill \\
\end{gathered}
\end{equation}
where the coupling strength in the continuum $g(\omega)$ is as
follows
\begin{equation}\label{B2}
    g(\omega)=\sqrt{\frac{\omega d}{4\pi\varepsilon_0\hbar v_g A}}
\end{equation}

Assume that the density of states is not flat. In this case, the
rule (\ref{B3}) is replaced with

\begin{equation}\label{B10}
\sum\limits_k {}  \to \frac{L} {{2\pi }}\int\limits_0^{ + \infty }
{d\omega }D(\omega)
\end{equation}
where $D(\omega)$ is the density of states per unit length and
unit frequency.

In order to proceed, we assume the density of states to have a
Loretzian profile (\ref{ds1}) which can be written as follows:
$D(\omega)=|d(\omega)|^2$ where

   \begin{equation}\label{B1}
d(\omega ) = \frac{\sqrt{\Gamma }} {{\omega  - \omega _0  +
i\Gamma  }},
   \end{equation}

In this case the discrete sum $\sum\limits_k {} g_k a_k^\dag$
transforms as follows

\begin{equation}\label{B11}
\begin{gathered}
  \sum\limits_k {} g_k a_k^\dag   \to \frac{L}
{{2\pi }}\int\limits_0^\infty  {d\omega |d(\omega )|^2 g_k } a_k^\dag   \hfill \\
   = \int\limits_0^\infty  {d\omega } \left( {\sqrt {\frac{L}
{{2\pi }}} \;g_k } \right)d(\omega )\left( {\sqrt {\frac{L}
{{2\pi }}} \;d^* (\omega )a_k^\dag  } \right)\hfill\\
 = \int\limits_0^\infty  {d\omega } \widetilde{g}(\omega )d(\omega )a_{}^\dag  (\omega ) \hfill \\
\end{gathered}
\end{equation}

where

\begin{equation}\label{B12}
    \widetilde{g}(\omega)=\sqrt{\frac{\omega d}{4\pi\varepsilon_0\hbar A}}
\end{equation}
and in analogy with (\ref{B4})

\begin{equation}\label{B13}
    a(\omega)^\dag=\frac{L}{2\pi}d^*(\omega)a_k^\dag
\end{equation}

Therefore, the derivation of expression (\ref{B11}) explains the
expression for Rabi frequency (\ref{Rabi_coh}).

\section{Derivation of equations (\ref{52}) and
(\ref{53b})}\label{C}

The Schrodinger equation $\widehat{Q} \Psi _{n  }  = q_{n_\alpha
} \Psi _{n } $ for the wavefunction (\ref{51b}) reads:
   \begin{equation}\label{A1}
\begin{gathered}
C_n \sqrt \Lambda  A^ +  \left| {g,n - 1  } \right\rangle_a
\\
 + C_n \int\limits_0^\infty  {d\omega } (\omega  - \Omega )a^ +  (\omega )a(\omega )\left| {e,n - 1  } \right\rangle_a
 \\
+ D_n \sqrt \Lambda  A\left| {e,n} \right\rangle_a  + D_n \int\limits_0^\infty  {d\omega } (\omega  - \Omega )a^ +  (\omega )a(\omega )\left| {g,n  } \right\rangle_a
\\
= q_{n_\alpha  } C_n \left| {e,n - 1  } \right\rangle_a  + q_{n_\alpha  } D_n \left| {g,n  } \right\rangle_a.
\end{gathered}
    \end{equation}

In order to proceed we use the following relations:
    \begin{equation}\label{A2}
\begin{gathered}
  \left[ {A,a_\alpha ^\dagger  } \right] = \frac{F}
{{\sqrt {\left\langle {n_\alpha  } \right\rangle } \sqrt \Lambda  }},
\\
  \left[ {\left( A \right)^{n'} \left( {a_\alpha ^\dagger  } \right)^n } \right] = \frac{{F^n }}
{{\left( {\sqrt {\left\langle {n_\alpha  } \right\rangle } } \right)^n \left( {\sqrt \Lambda  } \right)^n }}\delta _{n,n'},
\end{gathered}
   \end{equation}
where:
   \begin{equation}\label{AA}
F = \sum\limits_k {} \alpha _k^* g_k  \to \int\limits_0^\infty
{d\omega } d(\omega )\alpha ^* (\omega )g(\omega ),
   \end{equation}
   \begin{equation}\label{A3}
a(\omega )\left| {n  } \right\rangle_a  = \frac{{\alpha ^* (\omega
)}} {{\sqrt {\left\langle {n_\alpha  } \right\rangle } }}\sqrt n
\left| {n - 1  } \right\rangle_a,
   \end{equation}
   \begin{equation}\label{A4}
A\left| {n  } \right\rangle_a  = \frac{F} {{\sqrt {\left\langle
{n_\alpha  } \right\rangle } \sqrt \Lambda  }}\sqrt n \left| {n -
1  } \right\rangle_a.
   \end{equation}

With account of these relations equation (\ref{A1}) can be
rewritten as follows:
   \begin{equation}\label{A5}
\begin{gathered}
  C_n \sqrt \Lambda  A^\dagger  \left| {g,n - 1  } \right\rangle_a
\\
 + C_n \int\limits_0^\infty  {d\omega } (\omega  - \Omega )a^\dagger  (\omega )\frac{{\alpha ^* (\omega )}}
{{\sqrt {\left\langle {n_\alpha  } \right\rangle } }}\sqrt {n - 1} \left| {e,n - 2  } \right\rangle_a
\\
   + D_n \frac{F}
{{\sqrt {\left\langle {n_\alpha  } \right\rangle } }}\sqrt n
\left| {e,n - 1  } \right\rangle_a \\ + D_n \int\limits_0^\infty
{d\omega } (\omega  - \Omega )a^\dagger  (\omega )\frac{{\alpha ^*
(\omega )}}
{{\sqrt {\left\langle {n_\alpha  } \right\rangle } }}\sqrt n \left| {g,n - 1  } \right\rangle_a
\\
   = q_{n_\alpha  } C_n \left| {e,n - 1  } \right\rangle_a  + q_{n_\alpha  } D_n \left| {g,n  } \right\rangle_a  .
\end{gathered}
   \end{equation}

Multiplying the equation (\ref{A5}) on the left by $_a\left\langle
{n - 1,e  } \right|$ and then by
$_a\left\langle {n,g  } \right|$ we obtain two equations for
coefficients $C_n$ and $D_n$:
   \begin{equation}\label{A6}
\begin{gathered}
  C_n \int\limits_0^\infty  {d\omega } (\omega  - \Omega )\frac{{\alpha ^* (\omega )}}
{{\sqrt {\left\langle {n_\alpha  } \right\rangle } }}\sqrt {n - 1}\,\, _a\left\langle {n - 1,e  } \right|a^\dagger  (\omega )\left| {e,n - 2  } \right\rangle_a
\\
   + D_n \frac{F}
{{\sqrt {\left\langle {n_\alpha  } \right\rangle } }}\sqrt n  = q_{n_\alpha  } C_n,
\end{gathered}
   \end{equation}

\begin{equation}\label{A7}
\begin{gathered}
  C_n \int\limits_0^\infty  {d\omega } (\omega  - \Omega )\frac{{\alpha ^* (\omega )}}
{{\sqrt {\left\langle {n_\alpha  } \right\rangle } }}\sqrt {n - 1}\,\, _a\left\langle {n - 1,e } \right|a^\dagger  (\omega )\left| {e,n - 2  } \right\rangle_a
\\
   + D_n \frac{F}
{{\sqrt {\left\langle {n_\alpha  } \right\rangle } }}\sqrt n  = q_{n_\alpha  } D_n.
\end{gathered}
\end{equation}

Using equations (\ref{A3}), (\ref{A4}) we calculate the matrix
elements that enter equations (\ref{A6}), (\ref{A7}):
   \begin{equation}\label{A8}
_a\left\langle {n,g  } \right|A^\dagger  \left| {g,n - 1 }
\right\rangle_a  = \frac{{F^* }} {{\sqrt {\left\langle {n_\alpha }
\right\rangle } \sqrt \Lambda  }}\sqrt n,
   \end{equation}
   \begin{equation}\label{A9}
_a\left\langle {n - 1,e } \right|a^\dagger  (\omega )\left| {e,n - 2 }
\right\rangle_a  = \frac{{\alpha (\omega )}} {{\sqrt {\left\langle
{n_\alpha  } \right\rangle } }}\sqrt {n - 1},
   \end{equation}
   \begin{equation}\label{A10}
_a\left\langle {n,g  } \right|a^\dagger (\omega )\left| {g,n - 1 }
\right\rangle_a  = \frac{{\alpha (\omega )}} {{\sqrt {\left\langle
{n_\alpha  } \right\rangle } }}\sqrt {n}.
   \end{equation}

Substituting these matrix elements into equations (\ref{A6}),
(\ref{A7}) we obtain the equations (\ref{52}) and (\ref{53b}) from
the main text:
   \begin{equation}\label{A11}
C_n \left( {(n - 1)\Delta  - q_{n_\alpha  } } \right) + D_n
\frac{F} {{\sqrt {\left\langle {n_\alpha  } \right\rangle }
}}\sqrt n  = 0,
   \end{equation}
   \begin{equation}\label{A12}
C_n \frac{{F^* }} {{\sqrt {\left\langle {n_\alpha  } \right\rangle
} }}\sqrt n  + D_n (n\Delta  - q_{n_\alpha  } ) = 0.
   \end{equation}

\bibliography{Rabi_oscillations_4}

\begin{thebibliography}{99}


\bibitem{Brune1996}M. Brune, F. Schmidt-Kaler, A. Maali, J. Dreyer, E. Hagley, J. M.
Raimond, and S. Haroche, Quantum Rabi Oscillation: A Direct Test
of Field Quantization in a Cavity, Phys. Rev. Lett. \textbf{76},
1800 (1996).

\bibitem{Raim2001} J. M. Raimond, M. Brune, and S. Haroche, Manipulating quantum
entanglement with atoms and photons in a cavity, Rev. Mod. Phys.
\textbf{73}, 565 (2001).

\bibitem{Brune2004} M. Brune, Cavity Quantum Electrodynamics.
 Les Houches. Vol. \textbf{79}. Elsevier, 2004. 161-185.

\bibitem{Blais2004} A. Blais, R. S. Huang, A. Wallraff, S. M. Girvin, and R. J.
Schoelkopf, Cavity quantum electrodynamics for superconducting
electrical circuits: An architecture for quantum computation,
Phys. Rev. A \textbf{69}, 062320 (2004).

\bibitem{Blais2021} A. Blais, A.L. Grimsmo, S.M. Girvin, A. Wallraff, Circuit
quantum electrodynamics. Rev. Mod. Phys. \textbf{93}, 025005
(2021).

\bibitem{Wall2004} A. Wallraff, D. I. Schuster, A. Blais, L. Frunzio, R. S.
Huang, J. Majer, S. Kumar, S. M. Girvin, and R. J. Schoelkopf,
Strong coupling of a single photon to a superconducting qubit
using circuit quantum electrodynamics, Nature (London)
\textbf{431}, 162 (2004).

\bibitem{Wal2006} H. Walther, B. T. H. Varcoe, B.-G. Englert  and T. Becker,
Cavity quantum electrodynamics Rep. Prog. Phys. \textbf{69}, 1325
(2006).

\bibitem{Roy2017} D. Roy, C.M. Wilson, O. Firstenberg, Strongly interacting
photons in one-dimensional continuum. Rev. Mod. Phys. \textbf{89},
021001 (2017).

\bibitem{Sher2023} A.S. Sheremet, M.I. Petrov, I.V. Iorsh, A.V.
Poshakinskiy, and A.N. Poddubny, Waveguide quantum
electrodynamics: collective radiance and photon-photon
correlations. Rev. Mod. Phys. \textbf{95}, 015002 (2023).

\bibitem{Gu2017} X. Gu, A.F. Kockum, A. Miranowicz, Y.-X. Liu, F. Nori,
Microwave photonics with superconducting quantum circuits. Phys.
Rep. \textbf{718}, 1 (2017).

\bibitem{Mukh2024} D. Mukhopadhyay and J.-T. Shen,
Quantum multiphoton Rabi oscillations in waveguide QED. New J.
Phys. \textbf{26}, 103026 (2024).

\bibitem{Cohen2004}  C. Cohen-Tannoudji, J. Dupont-Roc, and G. Grynberg, Atom-
Photon Interactions: Basic Processes and Applications (Wiley- VCH,
Weinheim, 2004).

\bibitem{Green2025} Ya. S. Greenberga, O. A. Chuikin, A. G. Moiseev,
 A. A. Shtygashev, and O. V. Kibis, Quantum
correlations of the photon fields in a waveguide quantum
electrodynamics. Eur. Phys. J. B \textbf{98} 170 (2025).

\bibitem{Green2026} Ya. S. Greenberg , A. A. Shtygashev, and O. V. Kibis,
Decay of a transmon qubit in a broadband one-dimensional cavity.
Phys. Rev.A \textbf{113}, 042612 (2026).


\bibitem{Sult2025} A. Sultanov, E. Mutsenik, M. Schmelz, L. Kaczmarek,
G. Oelsner, U. H\"{u}bner, R. Stolz, and E. Il'ichev, Measuring
coherent dynamics of a superconducting qubit in an open waveguide.
Appl. Phys. Lett. 127, 042602 (2025).


\bibitem{Cheng2008}Y.-J. Cheng, Spatially dependent spontaneous emission and Rabi
oscillations of atoms in an open cavity with nonorthogonal
eigenmodes, Phys. Rev. A \textbf{77}, 033835 (2008).

 \bibitem{Scul1997} M. O. Scully and M. S. Zubairy, \textit{Quantum Optics}
(Cambridge University Press, Cambridge, England, 1997), Chap. 6.

\bibitem{Swain1972a} S. Swain, An exact analysis of spontaneous emission by
 a single two level atom in the rotating wave approximation. I. Analytic results.
J. Phys. A  Gen. Phys. \textbf{5}, 1587 (1972).

\bibitem{Swain1972b} S. Swain, An exact analysis of spontaneous emission
by a single two level atom in the rotating wave approximation. II.
Numerical results. J. Phys. A  Gen. Phys. \textbf{5}, 1601 (1972).

\bibitem{David1974} J. J. Yang, R. Davidson, and J. J. Kozak,
On the relaxation to quantum statistical equilibrium of the Wigner
Weisskopf atom in a one-dimensional radiation field. VI. Influence
of the coupling function on the dynamics. Journal of Mathematical
Physics \textbf{15}, 491 (1974).

\bibitem{Mandel1995}
L. Mandel and E. Wolf, \textit{Optical Coherence and Quantum
Optics} (Cambridge University Press, 1995), Chap. 10.

\bibitem{Blow1990} K. J. Blow, R. Loudon, and S. J. D. Phoenix, Continuum
fields in quantum optics. Phys. Rev. A \textbf{42}, 4102 (1990).

\bibitem{Dzs2016} D. Dzsotjan, B. Rousseaux, H. R. Jauslin,
G. Colas des Francs, C. Couteau, and S. Guerin, Mode-selective
quantization and multimodal effective models for spherically
layered systems. Phys. Rev. A \textbf{94}, 023818 (2016).

\bibitem{Dom2002}P. Domokos, P. Horak, and H. Ritsch, Quantum description of
light-pulse scattering on a single atom in waveguides. Phys. Rev.
A \textbf{65}, 033832 (2002).


\end{thebibliography}

\end{document}